\documentclass[onecolumn]{mn2e}
\input psfig.sty

\begin{document}

\title[Fundamental parameters of open clusters]{New fundamental parameters of the
Galactic open clusters Berkeley\,26, Czernik\,27, Melotte\,72, NGC\,2479 and BH\,37}

\author[Piatti et al.]{Andr\'es E. Piatti$^1$\thanks{E-mail:
andres@iafe.uba.ar (AEP); claria@oac.uncor.edu (JJC); aahumada@eso.org (AVA)} Juan J.
Clari\'a$^{2*}$ and Andrea V. Ahumada$^{2,3*}$\\
$^1$Instituto de Astronom\'{\i}a y F\'{\i}sica del Espacio, CC 67, Suc.
28, 1428, Ciudad de Buenos Aires, Argentina\\
$^{2}$Observatorio Astron\'omico, Universidad Nacional de C\'ordoba,
Laprida 854, 5000 C\'ordoba, Argentina\\
$^{3}$European Southern Observatory, Alonso de C\'ordova 3107, Santiago, Chile\\
}

\maketitle

\begin{abstract}

We have obtained CCD $UBVI_{KC}$ photometry down to V $\sim$ 21.0 for the open 
clusters Berkeley\,26, Czernik\,27, Melotte\,72, NGC\,2479 and BH\,37. The 
latter has never been studied before. Cluster stellar density profiles were 
obtained from star counts in appropriate-sized boxes distributed throughout the 
entire observed fields. Based on different measured indices, we estimate the ages 
of Berkeley\,26,  Melotte\,72 and NGC\,2479. On the other hand, we indicate 
possible solutions for the cluster fundamental parameters by matching theoretical 
isochrones which reasonably reproduce the main cluster features in their CMDs. In 
the case of NGC\,2479, the cluster $E(B-V)$ and $E(V-I)$ colour excesses and 
apparent distance modulus were estimated from the fit of the Zero-Age Main Sequence 
(ZAMS) to the colour-colour and colour-magnitude diagrams, respectively.
\end{abstract}


\section{Introduction}

Besides the intrinsic interest in open clusters (OCs) in their own right, it is generally 
accepted that these objects are fundamental landmarks to probe the Galactic disk 
properties (see, e.g., Friel 1995). They are among the very few Galactic objects for which 
meaningful distances can be derived over a large range, which makes them an essential tool 
to constrain Galactic evolution theories. They also make it possible to derive more 
accurate ages than are possible with other disk objects. Therefore, the study of the 
Galactic OC system proves very useful to clarify the many queries concerning the assessment of 
chemical abundance gradients in the disk (see, e.g., Twarog, Ashman \& Anthony-Twarog 1997; 
Chen, Hou \& Wang 2003), Galactic structure and evolution (e.g., Janes \& Adler 1982; 
Janes \& Phelps 1994), interactions between thin and thick disks (e.g., Sandage 1988), 
as well as theories of stellar formation and evolution (e.g., Meynet, Mermilliod \& 
Maeder 1993; Phelps \& Janes 1993).

The OC catalogue by Lyng\aa\ \shortcite{l87} includes 1700 entries. However, very little is 
known about many of them, except for their positions and approximate values of their 
angular sizes. At present, fewer than half of the known OCs have been studied in detail 
to derive their fundamental parameters. The current paper is part of a larger project 
aimed at looking into the formation and evolution of the Galactic disk by making use of a 
growing data base of photometric observations of as many OCs as possible. Thus, this study 
represents a further, intermediate step in a long-term programme devoted to obtain the 
fundamental parameters or to refine the quality of observationally determined properties for 
some unstudied or poorly studied OCs. 

$UBVI_{KC}$ photometry has proved to be a valuable tool to obtain the fundamental parameters 
of star clusters since information on cluster membership, distance, reddening, metallicity 
and age is obtained through the analysis of ($V$,$B-V$) and ($V$,$V-I$) Colour-Magnitude 
Diagrams (CMDs). In the year 2000, we carried out at Cerro Tololo Inter-American Observatory 
(CTIO, Chile) an observational program focused on still unstudied OCs at that time. We favoured 
the observation of OCs which were interesting not only because of the derivation of their basic 
parameters but also because of the possibility they offered of studying the morphology of 
their red giant evolutionary phases in relation to previous results (see, e.g., Mermilliod et 
al. 2001).

In this study, we report the results obtained from high-quality CCD $UBVI_{KC}$ photometry down 
to V $\approx$ 21.0 in the fields of the selected OCs Berkeley\,26, Czernik\,27, Melotte\,72, 
NGC\,2479 and BH\,37. These objects promised to be very interesting for their 
relatively old appearance due either to the observed stellar population in the CMDs or 
to their shapes and clustered nature. The basic parameters of the observed OCs are 
given in Table\,1, where the Trumpler class was taken from Archinal \& Hynes \shortcite{ah03}.
The last two columns list the total number of measured stars in this study and the 
inferred total number of cluster stars. The latter, together with the availability 
of a larger sample of data treated in the same way, will make it easier to establish a 
future calibration of the Trumpler richness class. All the selected clusters are located 
in the third Galactic quadrant near the Galactic plane ($\mid$$b$$\mid$ $\leq$ 6$\degr$). 
A brief description of these OCs, along with earlier photometric observations, is given 
below.

{\it Berkeley\,26}: Also known as Biurakan 12 \cite{i60} or C0647+058, this cluster seems to 
be a faint and probably old object in Monoceros. As indicated by its Trumpler class (III\,1m), it 
shows no strong central concentration but can be identified by its relatively dense population 
compared to that of the field stars (Fig. 1). Using 2-Micron All-Sky Survey (2MASS) data, Tadross 
\shortcite{t08} derived a heliocentric distance of 2.7 $\pm$ 0.1 kpc, $E(B-V)$ = 0.54 and 
an age of 600 Myr. These values, however, do not agree at all with those recently obtained by 
Hasegawa, Sakamoto \& Malasan \shortcite[hereafter HSM08]{hetal08} from CCD $VI$ photometry 
carried out with a 65 cm telescope. In fact, according to these authors, Berkeley\,26 is a very 
old (4.5 Gyr), highly reddened ($E(V-I)$ = 0.80) and very distant (d = 7.8 kpc) OC.

{\it Czernik\,27}: This is a relatively faint cluster first recognized in Monoceros by Czernik 
\shortcite{c66}. As indicated by its Trumpler class (III\,1p), Czernik\,27 (IAU designation 
C0700+064) is one of the most poorly defined objects of the present sample. It has a relatively 
small angular size of about 3$\arcmin$ \cite{l87}. Kim et al. \shortcite{kimetal05} and HSM08 
reported $BV$ and $VI$ CCD photometry in the cluster field, respectively. The cluster parameters 
determined in both cases, however, do not show a close correlation. Kim et al. 
\shortcite{kimetal05} 
found that Czernik\, 27 is a moderately reddened ($E(B-V)$ = 0.15) Hyades like age cluster 
located at 5.8 kpc from the Sun, while HSM08 concluded that this is a slightly reddened 
($E(B-V)$ $\approx$ 0.08) and older (1.1 Gyr) OC located at 4.3 kpc from the Sun.

{\it Melotte\,72}: According to Archinal \& Hynes \shortcite{ah03}, this object is the same 
as Collinder\,467 \cite{c31}. They described it as a small, compressed cluster, with its core 
about 3$\arcmin$ in diameter and with a 5$\arcmin$ long stream extending to the north 
between the two bright stars HD\,61277 ($V$ = 7.05, K5) and HD\,61401 ($V$ = 8.49, B9). The 
cluster lies about 1.3$\degr$ southwest of $\alpha$ Mon (Fig. 1). As far as we are 
aware, the only photometric study of this object was carried out by HSM08 using $VI$ CCD images. 
They found this object to be an intermediate-age cluster (1.6 Gyr) located at a distance of 3.2 kpc, 
with reddening $E(V-I)$ = 0.10.
 
{\it NGC\,2479}: This object, also referred to as Collinder\,167 \cite{c31}, C0752-175 or 
Trumpler\,8 \cite{t30}, is a relatively bright cluster in Puppis. As shown in Fig. 1, there are 
several bright stars in the cluster field, many of which seem to be foreground stars. 
Lyng\aa\ \shortcite{l87} reported an angular diameter of 11$\arcmin$ for NGC\,2479. Kharchenko 
et al. \shortcite{ketal05} presented a catalogue of astrophysical data for 520 Galactic OCs -among 
them NGC\,2479- which could be identified in their All-Sky Compiled Catalogue (ASCC-2.5). By 
applying homogeneous methods and algorithms, they determined basic parameters for their cluster 
sample. For NGC\,2479, they found the following results: $E(B-V)$ = 0.10, d = 1.2 kpc and $\sim$ 1 
Gyr. We should be cautious, however, when considering these findings since the limiting magnitude 
of the ASCC-2.5 is $V$ $\approx$ 12.5.

{\it BH\,37}: This is a detached, moderately rich and relatively faint OC (Fig. 1). As far as we 
know, no previous data exist for this compact object (IAU designation C0834-434) first recognized 
as an open cluster in Vela by van den Bergh \& Hagen \shortcite{vh75}.

The present photometric data are used to determine reddening, distance, age and metallicity of 
the selected OCs. The layout of the paper is as follows: Section 2 presents the observational 
material and the data reduction, whereas in Section 3 we determine the cluster centres and the 
stellar density radial profiles. In Section 4 we explain how to minimize the field star 
contamination in the CMDs. Section 5 deals with the determination of cluster fundamental 
parameters through the fitting of theoretical isochrones and with the comparison with 
previous results. Finally, Section 6 summarizes our findings and conclusions.

\section{Data collection and reduction}

We obtained images for the cluster sample in December 2000 with the $UBVI_{KC}$ filters and a 
2048$\times$2048 pixel Tektronix CCD attached to the CTIO 0.9 m telescope. The detector used  has a 
pixel size of 24 $\mu$m, producing a scale on the chip of 0.4\arcsec pixel$^{-1}$ (focal ratio 
f/13.5) and a 13.6$\arcmin$x13.6$\arcmin$ field of view. In order to standardize our photometry, we 
carried out observations of standard stars of the Selected Areas PG0231+051, 92 and 98 of Landolt 
\shortcite{l92}. 
By the end of each night, we had collected an average of 45 different measures of magnitude per 
filter for the selected standard star sample.

Table 2 shows the logbook of the observations with filters, exposure times, airmasses and 
seeing estimates. Observational setups, data reduction procedures, stellar point spread function 
photometry and transformation to the standard system, follow the same prescriptions described 
in detail in Piatti, Clari\'a \& Ahumada \shortcite{petal09}. The standard star photometry shows the 
mean square root deviation of the observations from the fits to be less than 0.015 mag, indicating 
that the nights were photometric. Once the standard magnitudes and colours were obtained, we 
produced a master table containing the average of $V$, $U-B$, $B-V$, and $V-I$, their 
errors $\sigma$($V$), $\sigma$($U-B$), $\sigma$($B-V$) and $\sigma$($V-I$) and the number of 
observations for each star, respectively. Tables 3 to 7 provide this information for Berkeley\,26, 
Czernik\,27, Melotte\,72, NGC\,2479 and BH\,37, respectively. Only a portion of these tables is 
shown here for guidance regarding their form and content. Tables 3 to 7 are shown in entirety 
in the online version of the journal. The deepest CCD images obtained for the cluster sample are 
shown in Fig. 1. In most cases the cluster region is only a small part of the observed frame, as 
indicated by the solid circles.

A simple inspection of Tables 3 to 7 shows that stars with three measures of $B-V$ and $V-I$ 
colours extend from the brightest limit down to $V$ = 19 mag and 20 mag, respectively. The 
stars with two measures of $B-V$ and $V-I$ colours cover $V$ ranges from 13.0 to 20.0 mag and 
from 13.0 to 21 mag, respectively. Finally, the stars with only one measure of $B-V$ and $V-I$ 
are fainter than $V$ = 18.0 and 19.0 mag, respectively, and they reach the photometric magnitude 
limits. According to these crude statistics, stars lying within the brightest 
$\sim$ 6 mags of our $\sim$ 9 mag range were measured two and three times. Therefore, they are 
the most appropriate ones to use to derive astrophysical information.
The behaviour of the photometric errors for the $V$ magnitude and $U-B$, $B-V$ and $V-I$ colours 
as a function of $V$ is shown in Table 8. Since those observed only once have practically 
no statistical weight, we decided to use all the stars. In addition, the knowledge of the 
behaviour of the photometric errors with the magnitude for these stars, allows us to rely on 
the accuracy of the morphology and position of the main cluster features in the CMDs. The 
resulting CMDs are drawn in Figs. 3, 4 and 5 which show, in general, broad sequences. It 
should be noticed that since we have measured just a handful of stars in the $U$ passband for 
Berkeley\,26, we only show the ($V$,$B-V$) and ($V$,$V-I$) CMDs for this cluster.

Kim et al. \shortcite{kimetal05} obtained $BV$ CCD photometry for stars in the field of
Czernik\,27. For 992 stars measured in common by Kim et al. \shortcite{kimetal05} and in this 
study, we derived $V_{\rm Kim}$ - $V_{\rm our}$ = 0.35 $\pm$ 0.07 mag, $(B-V)_{\rm Kim}$ - 
$(B-V)_{\rm our}$ = -0.05 $\pm$ 0.11 mag and $(V-I)_{\rm Kim}$ - 
$(V-I)_{\rm our}$ = 0.07 $\pm$ 0.08 mag, with a marginal dependence with the magnitude (see 
Fig. 2). Since Kim et al. \shortcite{kimetal05} also found $B$ and $V$ magnitude offsets when 
comparing their photometry for Berkeley\,29 with that of Kaluzny \shortcite{k94}, they decided 
to apply such offset to the Czernik\,27's photometry. This prevented us from using their data 
as a photometric reference. Kim et al. \shortcite{kimetal05} also obtained $BV$ data for only 
18 stars in the field of NGC\,2479 reaching a limiting magnitude of $V$ $\approx$ 12.5-13.0 mag, 
so that only a small portion of the cluster Main Sequence (MS) could be traced. When comparing 
their magnitudes and colours with those observed by us, we find $V_{\rm k05}$ - $V_{\rm our}$ = 
-0.03 $\pm$ 0.39 mag and $(B-V)_{\rm k05}$ - $(B-V)_{\rm our}$ = -0.02 $\pm$ 0.34 mag. Recently, 
HSM08 published CCD $VI$ photometry for Berkeley\,26, Czernik\,27 and Melotte\,72. Unfortunately, 
since the data are neither available in the WEBDA database \cite{mp03} nor upon request to 
the authors, we could not compare our photometry with theirs.

\section{Cluster dimensions and structure}

We first determined the location of the  clusters' centres in order to construct stellar 
density profiles. The coordinates of the  clusters' centres and their estimated uncertainties 
were determined, for each cluster, by fitting Gaussian distributions to the star counts 
in the $x$ and $y$ directions. The fits of the Gaussians were performed using the NGAUSSFIT 
routine in the STSDAS/IRAF package. We adopted a single Gaussian and fixed the constant to 
the corresponding background level and the linear terms to zero. The stars projected along the 
$x$ and $y$ directions were counted within intervals of 50 pixels. In addition, we checked 
that using spatial bins from 20 to 50 pixels or from 50 to 100 pixels does not lead to significant 
changes in the derived centres. We iterated the fitting procedure once on average, after 
eliminating a couple of discrepant points. Then, we determined the clusters' centres with a 
typical NGAUSSFIT standard deviation of $\pm$ 10 pixels. The centres of the Gaussians for 
Berkeley\,26, Czernik\,27, Melotte\,72, NGC\,2479 and BH\,37 were finally fixed at ($x_{c},y_{c}$) 
= (1170, 1300), (1320, 1270), (1370, 1210), (1215, 1250) and (1230, 1030) pixels, respectively. 

We constructed the clusters' radial profiles from star counts made in boxes of 50 pixels 
by 50 pixels, distributed throughout the whole field of each cluster. The chosen size of the box 
allowed us to sample, statistically, the stellar spatial distributions avoiding spurious effects 
caused mainly by the presence of localized groups of stars, rows or columns of stars. Thus, the 
number of stars per unit area, at a given radius $r$, can be directly calculated through the 
expression:

\begin{equation}
(n_{r+25} - n_{r-25})/((m_{r+25} - m_{r-25}) \times 50^2),
\end{equation}

\noindent where $n_j$ and $m_j$ represent the number of stars and boxes included in a circle of 
radius $j$, respectively. Note that this method does not necessarily require a complete circle of 
radius $r$ within the observed field to estimate the mean stellar density at such distance. This 
is important to consider since having a stellar density profile, which extends far away from the 
cluster centre, allows us to estimate the background level with higher precision. This is also 
helpful to measure the FWHM of the stellar density profile for it plays a significant role - from 
a stellar content point of view - in the construction of the cluster CMDs.

The resulting density profiles are shown in Fig. 6. The uncertainties estimated at various distances 
from the cluster centres follow Poisson statistics. Table 9 lists the estimated background 
levels, the radii at the FWHM ($r_{FWHM}$) and the field star contamination estimated in 
percentages. Note that the percentage of field stars is relatively high, which indicates a 
relatively small ratio between the number of each cluster's stars and the number of field stars. 
No cluster stands out clearly in its surrounding field within $r_{FWHM}$.

\section{Colour magnitude diagram cleaning}

Without a careful analysis of the observed sequences in the CMDs, one could come to the 
conclusion that they are in fact the clusters' MSs. However, all the CMDs present both cluster 
and field star MSs more or less superimposed. This means that we have observed both star 
clusters and their respective foreground fields affected by nearly similar reddenings, which 
makes it difficult to separate the fiducial cluster features and renders the analysis of 
the CMDs challenging.

To statistically clean the cluster CMDs from stars that can potentially belong to the 
foreground/background fields, we built star field CMDs using the stars located in the 
easternmost strip of the observed fields, i.e., $x$ $<$ 500 pixels and 0 $<$ $y$ (pixels) 
$<$ 2050 (see, Fig. 1). We separately treated the CMDs for $B-V$ and $V-I$. Using these field 
CMDs, we counted how many stars lie in different magnitude-colour bins with sizes 
[$\Delta$$V$, $\Delta$$(B-V)$=$\Delta$$(V-I)$] = (0.5,0.1) mag. We then subtracted from each 
cluster CMD the number of stars counted for each range of the field ($V$, $B-V$ or $V-I$) CMD, 
by removing those stars closer in magnitude and colour to the ones in the star field. Figs. 7, 
8 and 9 show the CMDs of the cluster surrounding field regions, while Figs. 10, 11 and 12 
show, with filled circles, the circular extracted CMDs which were obtained after cleaning 
them for field star  contamination. We show overplotted the CMDs directly obtained with all 
the measured stars (dots). When comparing observed and cleaned cluster CMDs, the differences 
in stellar composition became evident. Although the fiducial features of some clusters looked 
clearer, they appeared somewhat dispersed and scattered. This is mainly due to some 
unavoidable field interference. Other sources of dispersion such as photometric errors, 
differential internal cluster reddening, evolutionary effects and binarity can also 
account for such effect. In the subsequent analysis, we used the cleaned CMDs to estimate the 
cluster fundamental parameters. Note that we will only use the extracted ($V$,$V-I$) CMD of 
Berkeley\,26 since no star remains in its extracted ($V$,$B-V$) CMD.

\section{Estimates of the clusters fundamental parameters}

In order to estimate the ages of the observed clusters, we used the Morphological Age Index 
(MAI) defined by Janes \& Phelps \shortcite{jp94} on the basis of the $\delta$$V$, $\delta$$(B-V)1$ 
and $\delta$$(V-I)1$ indices of Phelps, Janes \& Montgomery \shortcite{phetal94} as well as 
the $\Delta$$V$ age index calibrated by  Carraro \& Chiosi \shortcite{cc94}. We also illustrate 
possible solutions for the fundamental cluster parameters by matching theoretical isochrones 
computed by Lejeune \& Schaerer \shortcite{ls01} to the observed CMDs. The previously known 
values of some cluster physical properties were used as reference to select the isochrones which 
best matched the CMDs.

{\it Berkeley\,26:} the region delimited by $V$ $<$ 18 and $V-I$ $>$ 1.4 would seem to contain 
cluster giants. We derived an age between 2 and 6 Gyr from the $\Delta$$V$ index, which 
agrees with the age resulting from the MAI (2.0 - 5.5 Gyr). If we use a 4 Gyr isochrone, which 
corresponds to the average age obtained from the MAI and the $\Delta$$V$ index, and we match it 
to the observed ($V$,$V-I$) CMD, we then get a fit consistent with the data. Indeed, the 
theoretical turnoff and subgiant branch magnitudes, the loci of the MS and of the red giant 
branch (RGB), and the slope of the RGB, all appear to be reasonably located with respect to the 
observed features. Using isochrones of 2 or 6 Gyr, we did not find successful fits that reproduced 
the observed $V-I$ distance between the MS turnoff and the bluest RGB point.

{\it Czernik\,27:} since it is not possible to obtain only one solution from the ZAMS fitting 
for the reddening-distance modulus pair, we assumed solar metallicity to evaluate which of both 
sets of published fundamental parameters (see Sect. 1) best resembles the fiducial features 
observed in the CMDs. We find the curvature and shape of the upper MS, the brightest magnitude 
of the MS and the bluest point of the turnoff, reasonably well fitted by the isochrones of 660 
Myr and 1.1 Gyr.

{\it Melotte\,72:} a Red Giant Clump (RGC) is visible at $V$ $\sim$ 13 mag and $B-V$ $\sim$ 
$V-I$ $\sim$ 1.0 mag. We derived an age between 0.4 and 1.0 Gyr from both the $\Delta$$V$ age 
index and the MAI. By using a solar metallicity 0.6 Gyr isochrone, it is possible to obtain a 
reasonable match to the cluster CMDs. The theoretical locus of the RGC, the brightest magnitude 
of the MS and the bluest point of the turnoff are the features which appear to be consistent 
with the data.

{\it NGC\,2479:} the cluster shows a very long star sequence and a compact RGC at $V$ $\sim$ 12 
and $U-B$ = $V-I$ $\sim$ 1.0 mag. Firstly, we derived the colour excesses from both colour-colour 
diagrams, by shifting the ZAMS along the directions of the corresponding reddening vectors 
\cite{s92}. Secondly, once the reddening effect was accounted for, we used the observed ZAMS 
($V$ $>$ 14 mag) to obtain the apparent distance modulus. The cluster age turned out to be in 
the 0.6 - 1.2 Gyr range from both the MAI and the  $\Delta$$V$ index. Moreover, by using a solar 
metallicity 1 Gyr isochrone, we achieved a good match to the fiducial features observed in the CMDs.

{\it BH\,37:} the cluster CMDs present a possible red giant branch and a reasonably well-defined 
evolved upper MS, particularly in the ($V$,$V-I$) CMD. As the cluster has not been studied in 
detail yet, we attempted a subjective isochrone match to the cluster 
CMDs.

Schlegel, Finkbeiner \& Davis's (1998, hereafter SFD) obtained full-sky maps 
from 100-μm dust emission. They found that in high Galactic latitude regions, the dust map 
correlates well with maps of H\,I emission. However, deviations are coherent in the sky 
and are especially conspicuous in regions of H\,I emission saturation towards denser 
clouds and in regions of formation of H$_2$ in molecular clouds \cite{petal03,petal08}. 
Even if the SFD's reddenings would not be exactly correct, they may still be valuable to 
compare with the reddenings derived here. We obtained $E(B-V)_{SFD}$ values of 0.61, 
0.19, 0.22, 0.16 and 2.19 mags for Berkeley\,26, Czernik\,27, Melotte\,72, NGC\,2479 and 
BH\,37, respectively. Since the $E(B-V)_{SFD}$ value for BH\,37 turned out to be 
more than double the one we estimated, we assumed that the $E(B-V)_{SFD}$ 
value must be saturated. It is worth considering that the ﬁve clusters lie further 
than 70 pc from the Galactic plane, with heights out of the Galactic plane of 0.17 
kpc, 0.49 kpc, 0.28 kpc, 0.14,kpc and -0.07, respectively.

Table 10 shows the values of the resulting fundamental parameters, while Figs. 13, 14 
and 15 indicate how the isochrones match the cluster features in the CMDs. Note 
that these values illustrate only possible solutions for the cluster fundamental parameters. 
Such solutions prove to be consistent with the data obtained. Fig. 13 shows that the MS of 
Berkeley\,26 is very broad, probably due to differential reddening and field star contamination. 
Although the three brighter and bluer stars than the turnoff are considered to be foreground 
stars, some of them could be blue stragglers. Similarly, some of the brighter and bluer 
stars than the MS turnoff of Czernik\,27 (Fig. 14) should also be considered blue straggler 
candidates. In order to improve the trace of the cluster features, it would be of great value 
to carry out deeper MS photometry and spectroscopic observations of the red giant branch. Note 
that the theoretically computed bluest stage, during the core He-burning phase, is redder than 
the observed RGC in the CMDs of Melotte\,72, a behaviour which has also been detected in other 
studies of Galactic and Magellanic Cloud clusters \cite[for example]{getal03,petal04a,petal04b}.


\subsection{Comparison with previous results}

{\it Berkeley\,26} : The cluster parameters found here are very different from those found by 
Tadross \shortcite{t08} from archival $JHK$ 2MASS photometry. By inspecting the CMDs obtained by 
Tadross to derive the cluster parameters (see his Fig. 6), we realized that he did not include 
upper MS and red giant branch stars. This error could have been due to the methods he 
employed for cleaning the CMDs and for selecting the cluster stars. In addition, the isochrone 
fit to the cluster CMDs does not resemble the cluster sequence at all. Our findings show a 
better agreement with the parameters recently determined by HSM08 from $VI$ CCD photometry. In 
fact, in  their study and in ours, the old and metal-poor character of Berkeley\, 26 is confirmed. 
Although the reddening derived in both studies agrees, within the errors, the heliocentric distance 
found by  HSM08 is larger than ours.   

{\it Czernik\,27} : Using Girardi et al.'s (2002) isochrones for solar metallicity content, 
Kim et al. \shortcite{kimetal05} estimated a cluster $E(B-V)$ reddening of 0.15, a distance from the 
Sun of 5.8 kpc, and an age of 600 Myr. HSM08 recently reported $VI$ CCD photometry in the cluster 
field. The parameters they found, however, do not show agreement, in general terms, with 
those derived by Kim et al. \shortcite{kimetal05}. Actually, according to HSM08, Czernik\,27 
seems to be reddened by scarcely $E(V-I)$ = 0.10, it is located closer to the Sun (4.3 kpc) and it is 
about 1.1 Gyr old. However, we find that both sets of parameters reasonably reproduce the 
current CMDs we obtained for the cluster (see Fig. 14).

{\it Melotte\,72} : As far as we are aware, the only detailed study of this cluster was carried 
out by  HSM08. Their heliocentric distance and metallicity are similar with our adopted 
values, their reddening ($E(V-I)$ = 0.25) and age (= 1.6 Gyr) being somewhat larger. However, when 
comparing their $(V,V-I)$ CMD for the central part of the cluster (see their Fig. 2) with our 
$(V,V-I)$ CMD (Fig. 14), we see that the turnoff of their selected isochrone is fainter that the 
turnoff we observed. We believe that the fit would have been much better if they had used a 
younger isochrone.

{\it NGC\,2479} : The catalogue by Kharchenko et al. \shortcite{ketal05}, based on the information 
provided by their ASCC-2.5, includes an analysis of 20 stars in the cluster field with $B,V$ 
magnitudes and proper motions in the Hipparcos system. Only a small portion of the cluster 
MS can be traced due to the limiting magnitude of these stars, which reaches $V$ $\approx$ 12.5-13.0 
mag. Based on a comprehensive analysis to determine membership, Kharchenko et al. showed 
that the 20 stars should be members according to the photometric data they used, although only 
four of them have proper motion membership probabilities higher than 80\%. The 20 stars do not 
define any clear MS in the ($V$,$B-V$) CMD, while the four stars with the highest proper motion 
membership probabilities are located between the cluster turnoff and the RGC. For comparison 
purposes, we have drawn these stars with open circles in Fig. 15. Surprisingly, all the cluster 
parameters determined by Kharchenko et al. (i.e., distance, age, reddening, core and cluster 
radii, etc.) coincide with our estimates.

\section{SUMMARY AND CONCLUSIONS}

New CCD $UBVI_{KC}$ photometry in the field of the open clusters Berkeley\,26, Czernik\,27, 
Melotte\,72, NGC\,2479 and BH\,37 is reported here. The analysis of the photometric data leads to 
the following main conclusions:

(i) Once the cluster centres were determined by fitting Gaussian distributions to the star 
counts in the $x$ and $y$ directions, radial density profiles were produced.

(ii) Cluster CMDs cleaned from field star contamination were built by statistically 
subtracting the number of stars counted in the field CMDs. Those stars closer in magnitude and 
colour to the ones in the respective star fields were thus removed.

(iii) Estimates of the cluster ages were obtained for Berkeley\,26, Melotte\,72 and 
NGC\,2479 from both the $\Delta$$V$ age index and the MAI. On the other hand, we outlined 
possible solutions for cluster fundamental parameters by matching theoretical isochrones,which 
reasonably reproduce the main cluster features in their CMDs. In the case of NGC\,2479, the 
$E(B-V)$ and $E(V-I)$ colour excesses and apparent distance modulus were estimated from the 
fit of the ZAMS to the colour-colour and magnitude-colour diagrams, respectively.




\section*{ACKNOWLEDGEMENTS}

We are gratefully indebted to the CTIO staff for their hospitality and support during the 
observations. We also thank referees Bruce Twarog and Kenneth Janes whose comments and 
suggestions have helped us to improve the manuscript. This work was partially supported by 
the Argentinian institutions CONICET, SECYT (Universidad Nacional de C\'ordoba) and Agencia 
Nacional de Promoci\'on Cient\'{\i}fica y Tecnol\'ogica (ANPCyT). This work is based 
on observations made at Cerro Tololo Inter-American Observatory, which is operated by AURA, 
Inc., under cooperative agreement with the National Science Foundation. This research 
also used the SIMBAD database, operated at CDS, Strasbourg, France; also the WEBDA database, 
operated at the Institute for Astronomy of the University of Vienna, and the NASA's 
Astrophysics Data.

\clearpage

\begin{table}
\caption{Basic parameters of the five open clusters.}
\begin{tabular}{@{}llccccccc}\hline
Cluster  & $\alpha_{\rm 2000}$  & $\delta_{\rm 2000}$  & {\it l}  &  $b$  & Trumpler Class  & Angular Diameter & Number of     & Number of inferred \\
         & (h m s)  & ($\degr$ $\arcmin$ $\arcsec$) & ($\degr$)  &  ($\degr$) & &  ($\arcmin$)                 & meaured stars & cluster stars \\
\hline
Berkeley\,26  &  6 50 18  & 5  45 00  & 207.69 &  2.35  & III\,1m  & 4.0   & 2226 & 97 \\
Czernik\,27   &  7 03 22  & 6  23 47  & 208.58 &  5.56  & III\,1p  & 3.0   & 2385 & 188 \\
Melotte\,72   &  7 38 24  & -10 41 00 & 227.84 &  5.36  & III\,1m  & 5.0   & 2770 & 107 \\
NGC\,2479     &  7 55 04  & -17 42 35 & 235.98 &  5.37  & III\,1m  & 11.0  & 2490 & 38 \\
BH\,37        &  8 35 49  & -43 37 00 & 262.35 & -1.78  & II\,1m   & 3.0   & 3000 & 65 \\
\hline
\end{tabular}
\end{table}
	 
\begin{flushleft}
\begin{table}
\caption{Observations log of selected clusters.}
\begin{tabular}{@{}lccccc}\hline
Cluster  & date & filter & exposure & airmass & seeing  \\
         &      &        &  (sec)   &         & ($\arcsec$)\\
\hline

Berkeley\,26 &  Dec. 29, 2000 & $U$ & 60 & 1.24 & 1.5 \\
             &                & $U$ &420 & 1.24 & 1.4 \\
             &                & $V$ & 20 & 1.24 & 1.2 \\
             &                & $V$ & 60 & 1.24 & 1.4 \\
             &                & $V$ &200 & 1.23 & 1.4 \\
             &                & $I$ & 10 & 1.23 & 1.1 \\
             &                & $I$ & 90 & 1.23 & 1.4 \\
             &                & $I$ & 90 & 1.24 & 1.2 \\

Czernik\,27  &  Dec. 23, 2000 & $B$ & 20 & 1.30 & 1.8 \\
             &                & $B$ & 60 & 1.30 & 2.0 \\
             &                & $B$ &360 & 1.30 & 2.0 \\
             &                & $V$ & 20 & 1.32 & 1.6 \\
             &                & $V$ & 60 & 1.32 & 2.0 \\
             &                & $V$ &200 & 1.32 & 2.0 \\
             &                & $I$ & 10 & 1.28 & 1.1 \\
             &                & $I$ & 90 & 1.28 & 1.3 \\

Melotte\,72  &  Dec. 24, 2000 & $B$ & 20 & 1.08 & 1.8 \\
             &                & $B$ & 60 & 1.08 & 1.9 \\
             &                & $B$ &360 & 1.08 & 1.9 \\
             &                & $V$ & 20 & 1.09 & 1.3 \\
             &                & $V$ & 60 & 1.09 & 1.6 \\
             &                & $V$ &360 & 1.08 & 1.6 \\
             &                & $I$ & 10 & 1.07 & 1.3 \\
             &                & $I$ & 90 & 1.07 & 1.3 \\

NGC\,2479    &  Dec. 29, 2000 & $U$ & 20 & 1.30 & 1.8 \\
             &                & $U$ &540 & 1.04 & 1.5 \\
             &                & $B$ & 20 & 1.03 & 1.6 \\
             &                & $B$ & 60 & 1.03 & 1.6 \\
             &                & $B$ &360 & 1.03 & 1.5 \\
             &                & $V$ & 20 & 1.02 & 2.0 \\
             &                & $V$ & 60 & 1.02 & 1.4 \\
             &                & $V$ &200 & 1.03 & 1.3 \\
             &                & $I$ & 10 & 1.02 & 1.1 \\
             &                & $I$ & 90 & 1.02 & 1.3 \\

BH\,37       &  Dec. 24, 2000 & $B$ & 20 & 1.03 & 1.3 \\
             &                & $B$ & 60 & 1.03 & 1.4 \\
             &                & $B$ &360 & 1.03 & 1.5 \\
             &                & $V$ & 20 & 1.03 & 1.3 \\
             &                & $V$ & 60 & 1.03 & 1.4 \\
             &                & $V$ &360 & 1.03 & 1.4 \\
             &                & $I$ & 10 & 1.03 & 1.2 \\
             &                & $I$ & 90 & 1.03 & 1.3 \\
\hline
\end{tabular}
\end{table}
\end{flushleft}

\begin{flushleft}
\begin{table}
\caption{CCD $VI$ data of stars in the field of Berkeley\,26.}
\begin{tabular}{@{}lcccccccc}\hline
ID & $x$ & $y$ & $V$ & $\sigma$$(V)$ & n$_V$ & $V-I$ & $\sigma$$(V-I)$ & n$_{VI}$\\
     & (pix) & (pix) & (mag) & (mag) &  & (mag) & (mag) & \\\hline
     & ...   & ...   & ...   & ...   & .&...    & ...   & \\ 
    70 &1400.643 &  61.032 & 19.420 &0.067&  1&  1.345 &0.084 & 1\\
    71 &1122.117 &  61.227 & 18.361 &0.031&  3&  1.565 &0.072 & 2\\
    72 &1370.823 &  62.633 & 18.533 &0.053&  2&  1.332 &0.004 & 2\\
   ...& ...     & ...     & ...   & ...   &  .&... & ...   & . \\
   ...& ...     & ...     & ...   & ...   & .&... & ...   & . \\
   ...& ...     & ...     & ...   & ...   & .&... & ...   & . \\
\hline
\end{tabular}
\medskip

\noindent {\sc NOTE:} ($x$,$y$) coordinates correspond to the reference 
system of Fig. 1. Magnitude and colour errors are the standard deviations of 
the mean or the observed photometric errors for stars with only one 
measurement.

\end{table}
\end{flushleft}

\begin{flushleft}
\begin{table}
\caption{CCD $BVI$ data of stars in the field of Czernik\,27.}
\begin{tabular}{@{}lccccccccccc}\hline
ID & $x$ & $y$ & $V$ & $\sigma$$(V)$ & n$_V$ &  $B-V$ & $\sigma$$(B-V)$ & n$_{BV}$ & $V-I$ & $\sigma$$(V-I)$ & n$_{VI}$\\
     & (pix) & (pix) & (mag) & (mag) &  &   (mag) & (mag) &  & (mag) & (mag) & \\\hline
   ...& ...     & ...     & ...   & ...   & .&...& ...   & .&... & ...   & . \\ 
    88 & 399.518 &  68.806 & 15.750 &0.044 & 3 &    0.714& 0.027 & 2 &  0.816 &0.067 & 3\\
    89 & 419.049 &  70.068 & 16.170 &0.062 & 3 &    0.643& 0.029 & 2 &  0.745 &0.073 & 3\\
    90 &1737.327 &  71.037 & 20.173 &0.049 & 1 &   99.999& 9.999 & 0 &  1.255 &0.095 & 1\\
   ...& ...     & ...     & ...   & ...   & .&...& ...   & .&... & ...   & . \\
   ...& ...     & ...     & ...   & ...   & .&...& ...   & .&... & ...   & . \\
   ...& ...     & ...     & ...   & ...   & .&...& ...   & .&... & ...   & . \\
\hline
\end{tabular}
\medskip

\noindent {\sc NOTE:} ($x$,$y$) coordinates correspond to the reference 
system of Fig. 1. Magnitude and colour errors are the standard deviations of 
the mean or the observed photometric errors for stars with only one 
measurement.
\end{table}
\end{flushleft}

\begin{flushleft}
\begin{table}
\caption{CCD $BVI$ data of stars in the field of Melotte\,72.}
\begin{tabular}{@{}lccccccccccc}\hline
ID & $x$ & $y$ & $V$ & $\sigma$$(V)$ & n$_V$ &  $B-V$ & $\sigma$$(B-V)$ & n$_{BV}$ & $V-I$ & $\sigma$$(V-I)$ & n$_{VI}$\\
     & (pix) & (pix) & (mag) & (mag) &  &   (mag) & (mag) &  & (mag) & (mag) & \\\hline
   ...& ...     & ...     & ...   & ...   & .&...& ...   & .&... & ...   & . \\ 
   402 &1684.695&  300.844&  20.853 &0.145 & 2 &   1.727 &0.195 & 1 &  1.831 &0.144 & 2\\
   403 &1751.192&  302.558&  20.173 &0.033 & 2 &   1.152 &0.066 & 1 &  1.257 &0.032 & 2\\
   404 &1759.010&  304.554&  19.237 &0.001 & 3 &   0.891 &0.012 & 2 &  0.995 &0.024 & 3\\
   ...& ...     & ...     & ...   & ...   & .&...& ...   & .&... & ...   & . \\
   ...& ...     & ...     & ...   & ...   & .&...& ...   & .&... & ...   & . \\
   ...& ...     & ...     & ...   & ...   & .&...& ...   & .&... & ...   & . \\
\hline
\end{tabular}
\medskip

\noindent {\sc NOTE:} ($x$,$y$) coordinates correspond to the reference 
system of Fig. 1. Magnitude and colour errors are the standard deviations of 
the mean or the observed photometric errors for stars with only one 
measurement.

\end{table}
\end{flushleft}

\begin{flushleft}
\begin{table}
\caption{CCD $UBVI$ data of stars in the field of NGC\,2479.}
\begin{tabular}{@{}lcccccccccccccc}\hline
ID & $x$ & $y$ & $V$ & $\sigma$$(V)$ & n$_V$ & $U-B$ & $\sigma$$(U-B)$ & n$_{UB}$ & $B-V$ & $\sigma$$(B-V)$ & n$_{BV}$ & $V-I$ & $\sigma$$(V-I)$ & n$_{VI}$\\
     & (pix) & (pix) & (mag) & (mag) &  & (mag) & (mag) & &  (mag) & (mag) &  & (mag) & (mag) & \\\hline
   ...& ...     & ...     & ...   & ...   & .&...    & ... &.&...& ...   & .&... & ...   & . \\ 
    11 & 988.318&  390.190&  18.199& 0.010&  3&   1.414& 0.106&  1&   0.849& 0.022&  1&   1.335& 0.009&  3\\
    12 &1700.800& 1030.361&  18.505& 0.024&  3&   0.674& 0.014&  2&   0.645& 0.003&  2&   0.990& 0.023&  3\\
    13 & 158.117& 1768.943&  16.464& 0.011&  3&   0.105& 0.002&  2&   0.446& 0.007&  3&   0.658& 0.011&  3\\
   ...& ...     & ...     & ...   & ...   & .&...    & ... &.&...& ...   & .&... & ...   & . \\
   ...& ...     & ...     & ...   & ...   & .&...    & ... &.&...& ...   & .&... & ...   & . \\
   ...& ...     & ...     & ...   & ...   & .&...    & ... &.&...& ...   & .&... & ...   & . \\
\hline
\end{tabular}
\medskip

\noindent {\sc NOTE:} ($x$,$y$) coordinates correspond to the reference 
system of Fig. 2. Magnitude and colour errors are the standard deviations of 
the mean or the observed photometric errors for stars with only one 
measurement.

\end{table}
\end{flushleft}

\begin{flushleft}
\begin{table}
\caption{CCD $BVI$ data of stars in the field of BH\,37.}
\begin{tabular}{@{}lccccccccccc}\hline
ID & $x$ & $y$ & $V$ & $\sigma$$(V)$ & n$_V$ &  $B-V$ & $\sigma$$(B-V)$ & n$_{BV}$ & $V-I$ & $\sigma$$(V-I)$ & n$_{VI}$\\
     & (pix) & (pix) & (mag) & (mag) &  &   (mag) & (mag) &  & (mag) & (mag) & \\\hline
   ...& ...     & ...     & ...   & ...   & .&...& ...   & .&... & ...   & . \\ 
142 &1227.468&  100.013&  17.589& 0.028&  2&   1.250 &0.036 & 2&   1.523& 0.039 & 2\\
143 &1455.322&  100.024&  19.965& 0.011&  2&   1.374 &0.080 & 1&   1.646& 0.125 & 2\\
144 &1476.955&  100.562&  21.489& 0.098&  1&   99.999& 9.999&  0 &  2.185& 0.131&  1\\
   ...& ...     & ...     & ...   & ...   & .&...& ...   & .&... & ...   & . \\
   ...& ...     & ...     & ...   & ...   & .&...& ...   & .&... & ...   & . \\
   ...& ...     & ...     & ...   & ...   & .&...& ...   & .&... & ...   & . \\
\hline
\end{tabular}
\medskip

\noindent {\sc NOTE:} ($x$,$y$) coordinates correspond to the reference 
system of Fig. 1. Magnitude and colour errors are the standard deviations of 
the mean or the observed photometric errors for stars with only one 
measurement.

\end{table}
\end{flushleft}

\begin{table}
\caption{Magnitude and colour photometric errors as a function of V.}
\begin{tabular}{@{}ccccc}\hline
$\Delta$$V$ & $\sigma$$(V)$ & $\sigma$$(U-B)$ & $\sigma$$(B-V)$ & 
$\sigma$$(V-I)$ \\
   (mag)    &    (mag)    &     (mag)     &     (mag)     &    (mag)   \\
\hline
11-12 & $<$ 0.01 & 0.01 & $<$ 0.01 & $<$ 0.01 \\
12-13 &$<$ 0.01 & 0.02 & 0.01      & 0.01\\
13-14 &$<$ 0.01 & 0.02 & 0.01      & 0.01\\
14-15 & 0.01   &  0.03 & 0.01      & 0.02\\
15-16 & 0.01   &  0.04 & 0.01      & 0.02\\
16-17 & 0.02   &  0.04 & 0.02      & 0.02\\
17-18 & 0.02   &  0.05 & 0.02      & 0.03\\
18-19 & 0.03   &  0.06 & 0.03      & 0.03\\
19-20 & 0.03   &  0.07 & 0.05      & 0.05\\
20-21 & 0.05   &  0.10 & 0.10      & 0.10\\
21-22 & 0.10   &  ---  & ---       & 0.15\\
\hline
\end{tabular}
\end{table}

\begin{table}
\caption{Cluster sizes and field contamination.}
\begin{tabular}{@{}lccc}\hline
Name &   Background$^a$  & $r$$_{FWHM}$  & Field contamination (\%) \\
     &     (counts)      &  (px)         &  $r$ $<$ $r$$_{FWHM}$ \\
\hline
Berkeley\,26 & 3.6   &  140 $\pm$ 15 &    19  \\
Czernik\,27  & 4.3   &  125 $\pm$ 15 &    41  \\
Melotte\,72  & 5.0   &  200 $\pm$ 15 &    57  \\
NGC\,2479    & 5.0   &  150 $\pm$ 15 &    63  \\
BH\,37       & 5.0   &  180 $\pm$ 15 &    43  \\

\hline
\end{tabular}
\medskip

\noindent $^a$ Normalized to a circular area of radius 50 pixels.

\end{table}

\begin{table}
\caption{Possible solutions for the fundamental parameters of the selected clusters.}
\begin{tabular}{@{}lccccc}\hline
                   & Berkeley\,26    & Czernik\,27  & Melotte\,72             & NGC\,2479               & BH\,37 \\
\hline
$E(B-V)$ (mag)     & 0.75      & 0.08 / 0.15  & 0.20         & 0.05         & 1.05 \\
$E(V-I)$ (mag)     & 0.95      & 0.10 / 0.20  & 0.25         & 0.07         & 1.30 \\
$m-M_V$  (mag)     & 15.50     & 13.4 / 14.3  & 13.00        & 11.00        & 15.25 \\
Age (Gyr)          &  4        & 0.7 / 1.1    & 0.60         & 1.00         &  0.70 \\
{\rm [Fe/H]} (dex) & -0.7      & 0.0          &  0.0         & 0.0          &  0.0 \\

\hline
\end{tabular}
\end{table}

\clearpage

\begin{figure}
\centerline{\psfig{figure=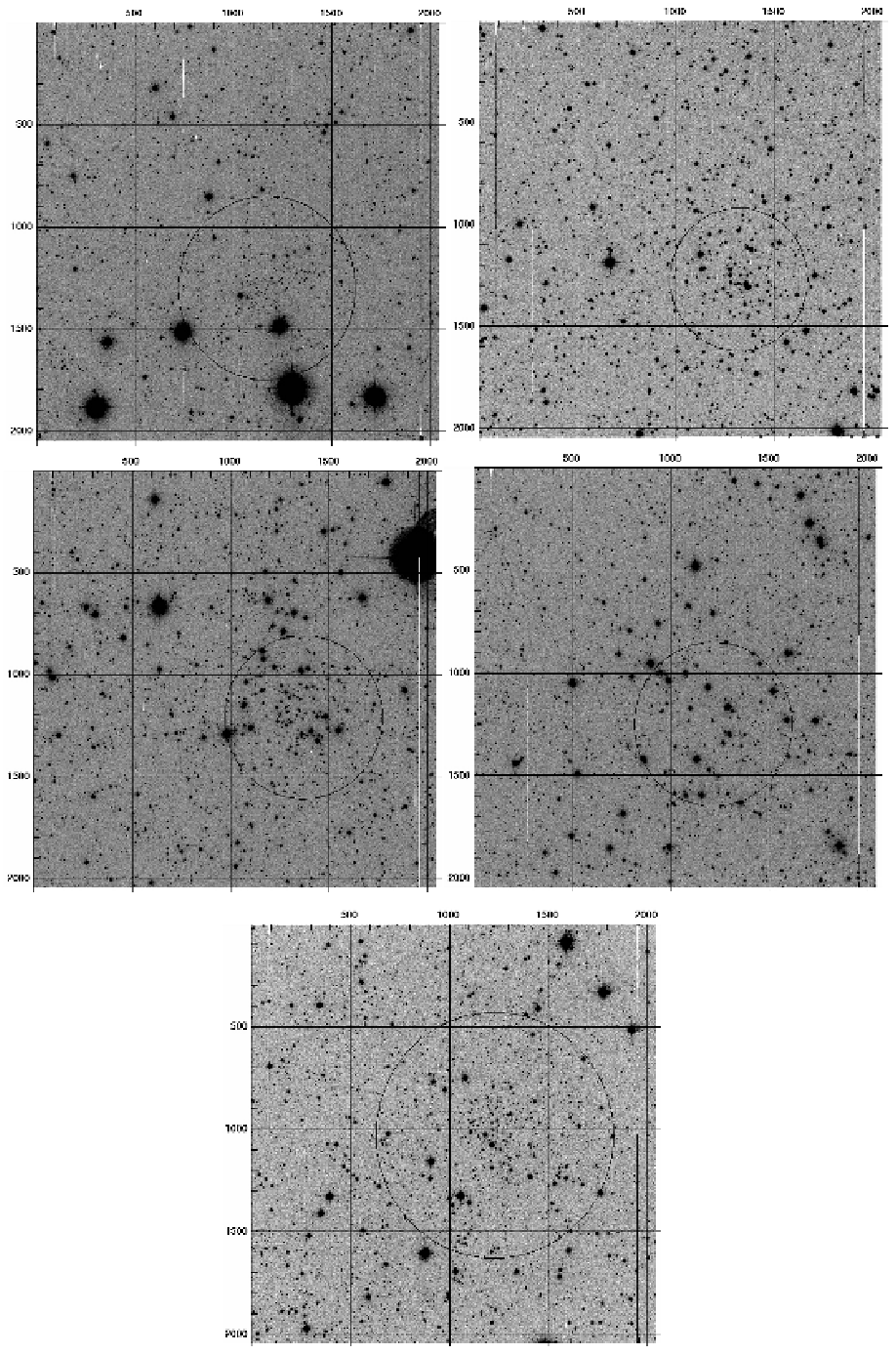,width=168mm}}
\caption{Deepest CCD images obtained : 90 s $I$ for Berkeley\,26 
({\it top left}); 360 s $B$ for Czernik\,27 ({\it top right});
90 s $I$ for Melotte\,72 ({\it middle left}), 
90 s $I$ for NGC\,2479 ({\it middle right}), and 200 s $V$ for 
BH\,37 ({\it bottom}). The circles encompass the cluster regions. North is up and East is to the left. Coordinates are given in pixels.}
\label{fig1}
\end{figure}

\begin{figure}
\centerline{\psfig{figure=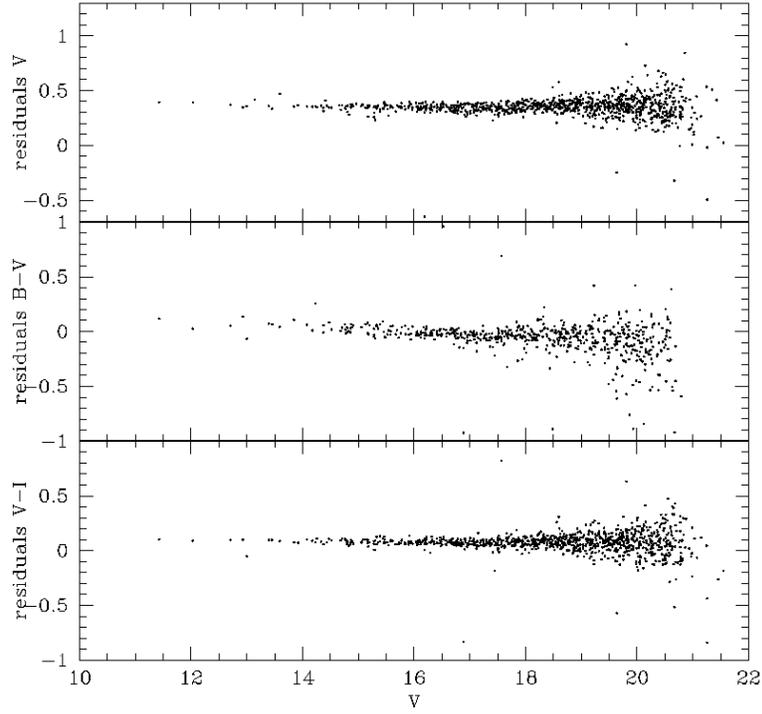}}
\caption{Comparison between our photometry for Czernik\,27 and that of Kim et al. (2005).}
\label{fig2}
\end{figure}

\begin{figure}
\centerline{\psfig{figure=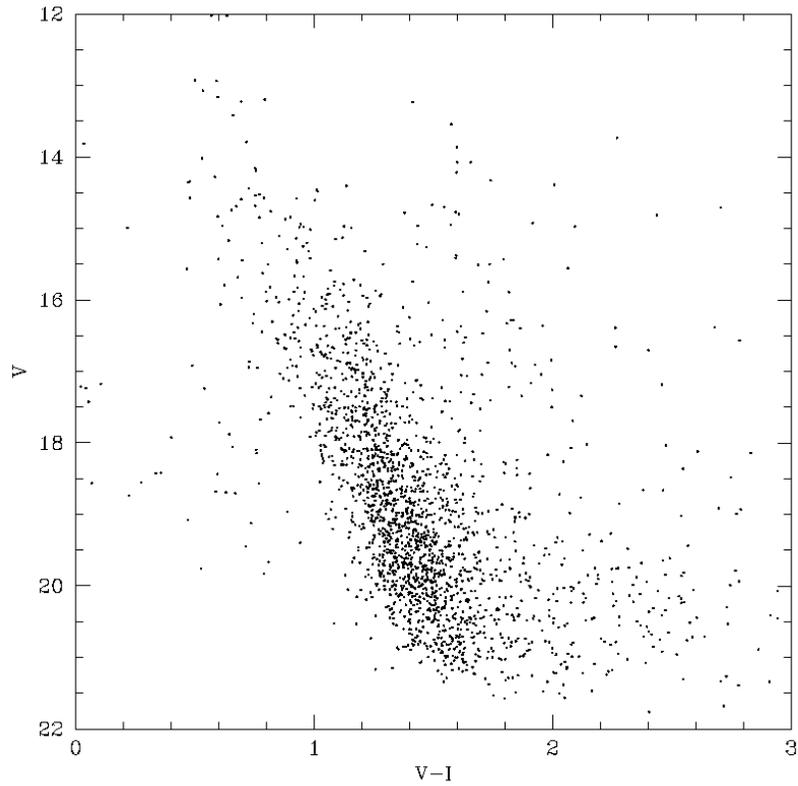}}
\caption{The ($V,V-I$) diagram for the stars measured in the field of Berkeley\,26.}
\label{fig3}
\end{figure}

\begin{figure}
\centerline{\psfig{figure=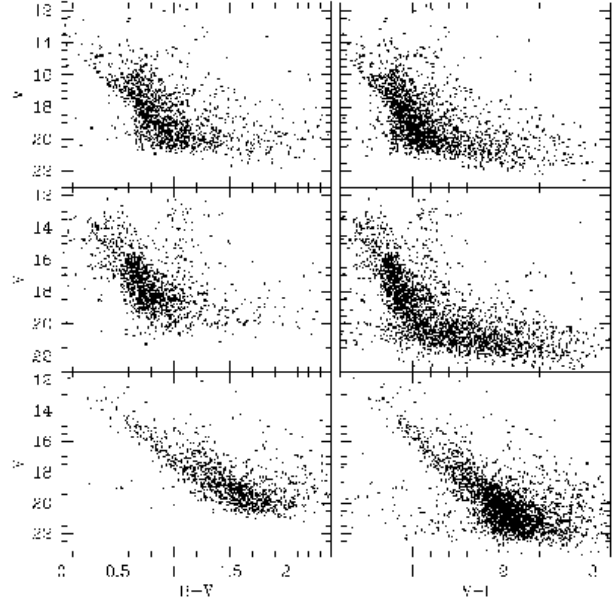}}
\caption{The ($V,B-V$) and ($V,V-I$) diagrams for the stars measured in the fields of 
Czernik\,27 {\it (top)}, Melotte\,72 ({\it middle}) and BH\,37 ({\it bottom}).}
\label{fig4}
\end{figure}

\begin{figure}
\centerline{\psfig{figure=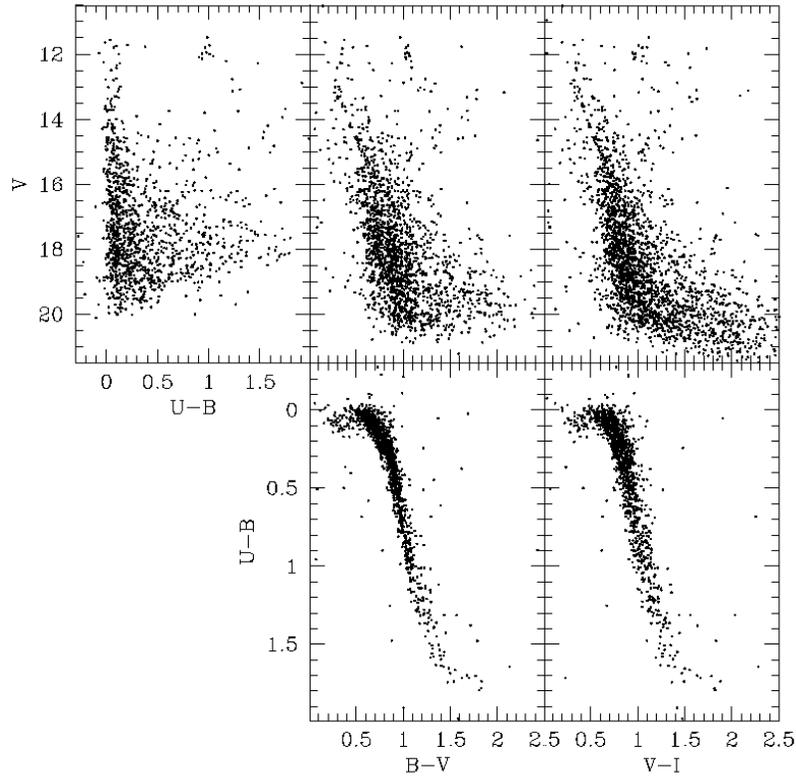}}
\caption{The ($V,U-B$), ($V,B-V$), and ($V,V-I$) diagrams ({\it top}), and
($U-B,B-V$) and ($B-V,V-I$) diagrams ({\it bottom}) for the stars measured in the field of NGC\,2479.}
\label{fig5}
\end{figure}

\begin{figure}
\centerline{\psfig{figure=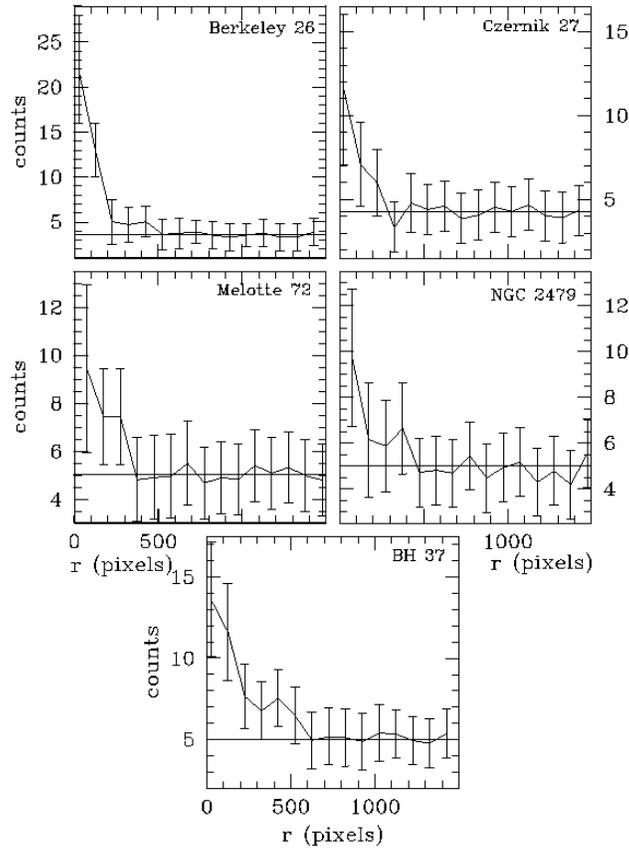}}
\caption{Cluster stellar density radial profiles normalized to a circular
area of 50 pixel radius. The horizontal lines represent the measured background levels (see Section 3 for details).}
\label{fig6}
\end{figure}

\begin{figure}
\centerline{\psfig{figure=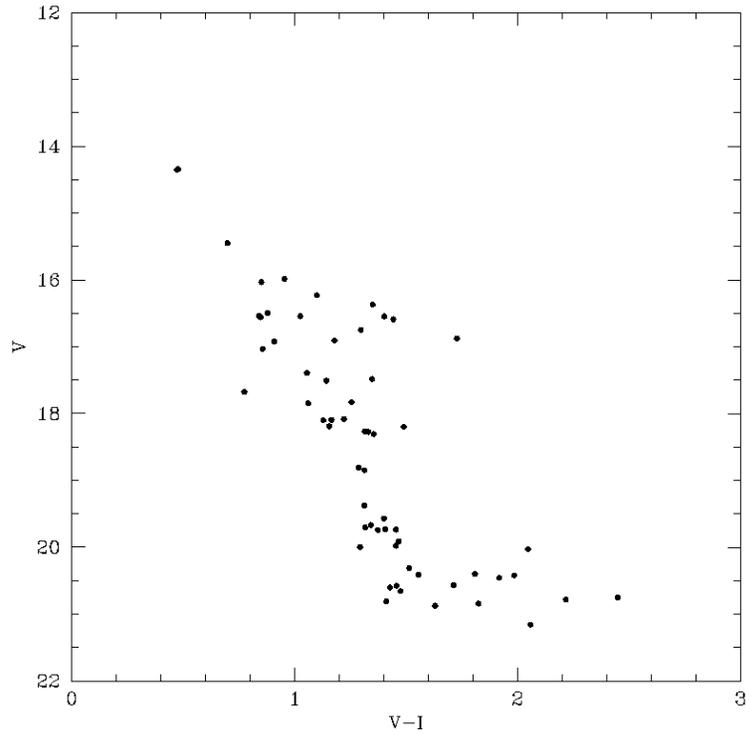}}
\caption{The ($V,V-I$) diagram for the surrounding field region of Berkeley\,26,
normalized to a circular area of radius 200 pixels for comparison with Fig. 10.}
\label{fig7}
\end{figure}

\begin{figure}
\centerline{\psfig{figure=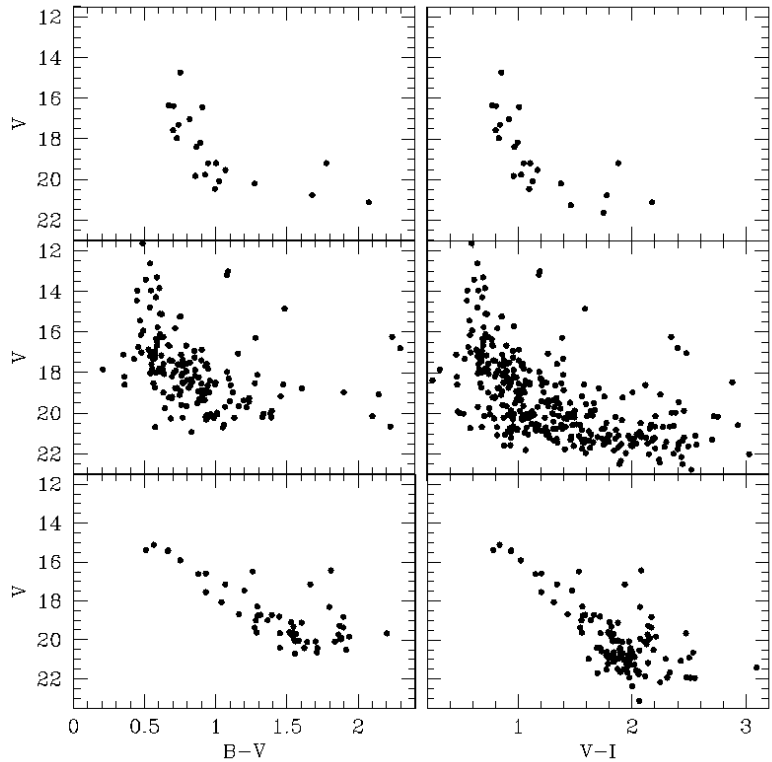}}
\caption{The ($V,B-V$) and ($V,V-I$) diagrams for the surrounding field regions of 
Czernik\,27 {\it (top)}, Melotte\,72 ({\it middle}) and BH\,37 ({\it bottom}),
normalized to a circular area of radius 200 pixels for comparison with Fig. 11.}
\label{fig8}
\end{figure}

\begin{figure}
\centerline{\psfig{figure=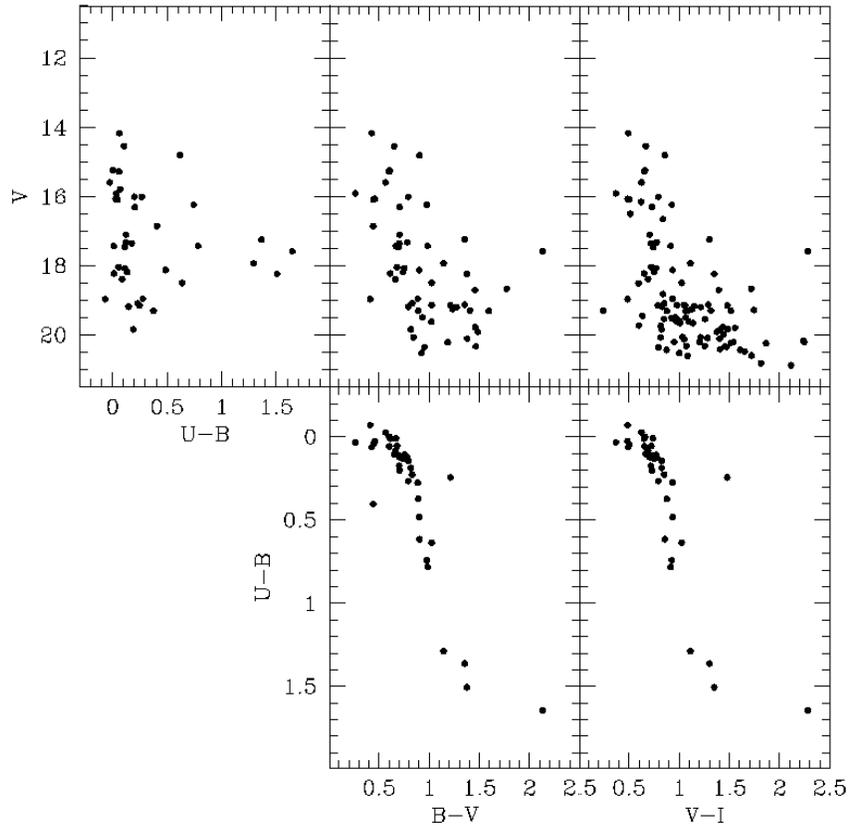}}
\caption{The ($V,U-B$), ($V,B-V$), and ($V,V-I$) diagrams ({\it top}), and
($U-B,B-V$) and ($B-V,V-I$) diagrams ({\it bottom}) for the surrounding field region
of NGC\,2479, normalized to a circular area of radius 200 pixels for comparison with Fig. 12.}
\label{fig9}
\end{figure}

\begin{figure}
\centerline{\psfig{figure=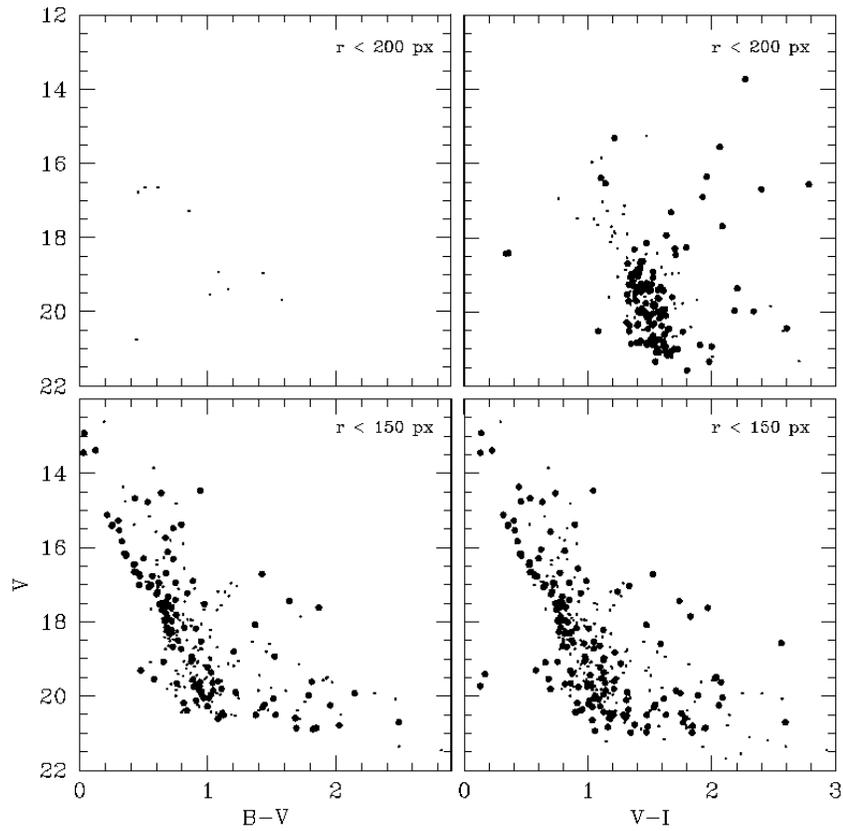}}
\caption{The circular extracted ($V,B-V$) and ($V,V-I$) diagrams for the clusters (dots) Berkeley\,26 
({\it top}) and Czernik\,27 ({\it bottom}), compared with those statistically cleaned from field
star contamination (filled circles).}
\label{fig10}
\end{figure}

\begin{figure}
\centerline{\psfig{figure=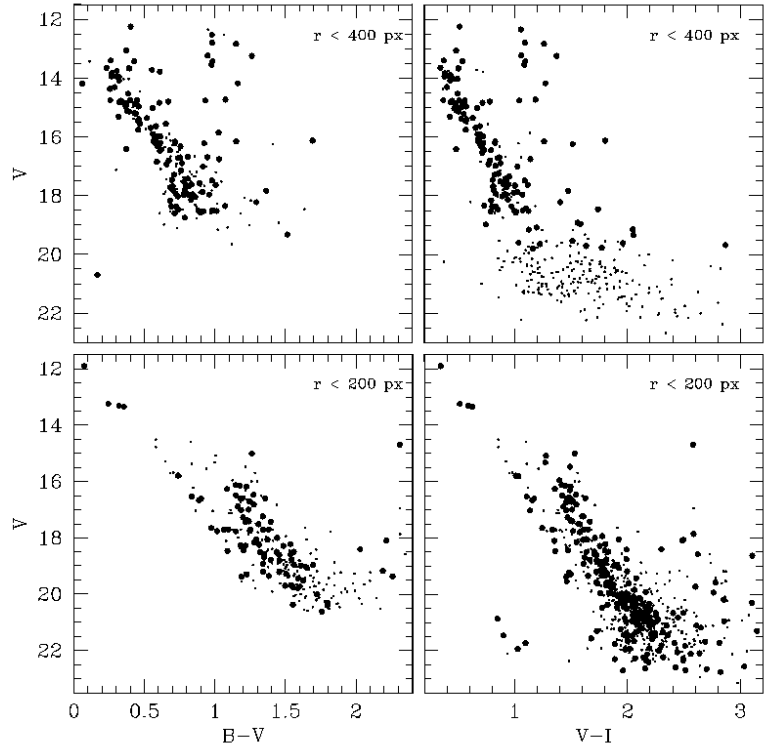}}
\caption{The circular extracted ($V,B-V$) and ($V,V-I$) diagrams for the clusters (dots) Melotte\,72 
({\it top}) and BH\,37 ({\it bottom}), compared with those statistically cleaned from field
star contamination (filled circles).}
\label{fig11}
\end{figure}

\begin{figure}
\centerline{\psfig{figure=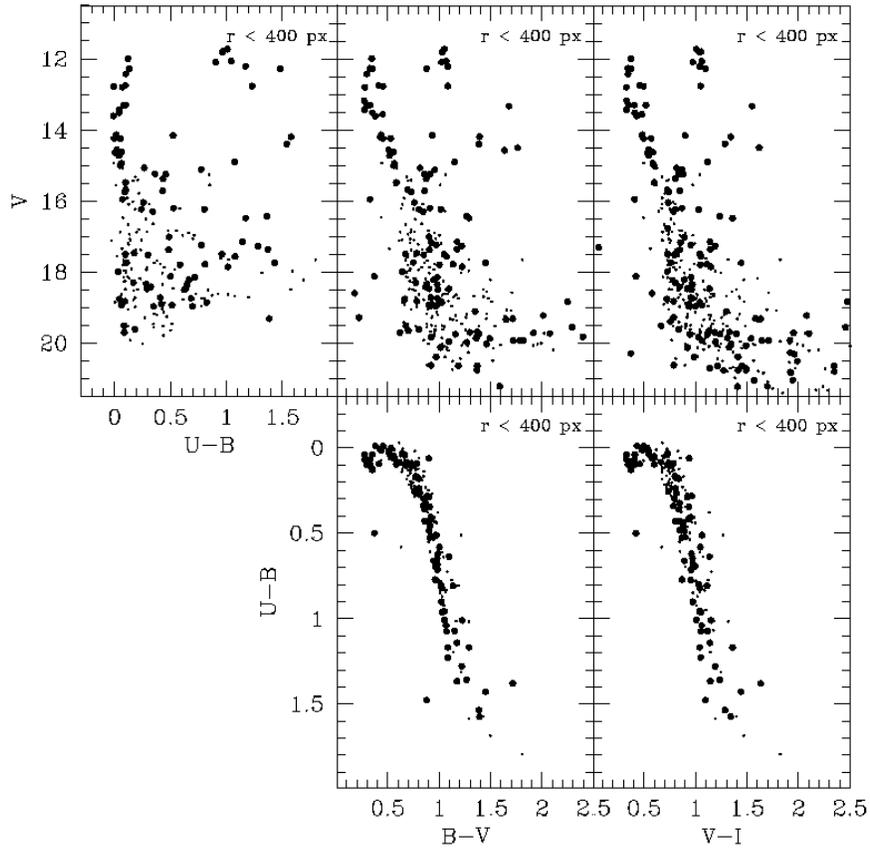}}
\caption{ The circular extracted ($V,U-B$), ($V,B-V$), and ($V,V-I$) diagrams ({\it top}), and circular extracted ($U-B,B-V$) and ($B-V,V-I$) diagrams ({\it bottom}) for the cluster NGC\,2479 (dots), compared with those statistically cleaned from field star contamination (filled circles).}
\label{fig12}
\end{figure}

\begin{figure}
\centerline{\psfig{figure=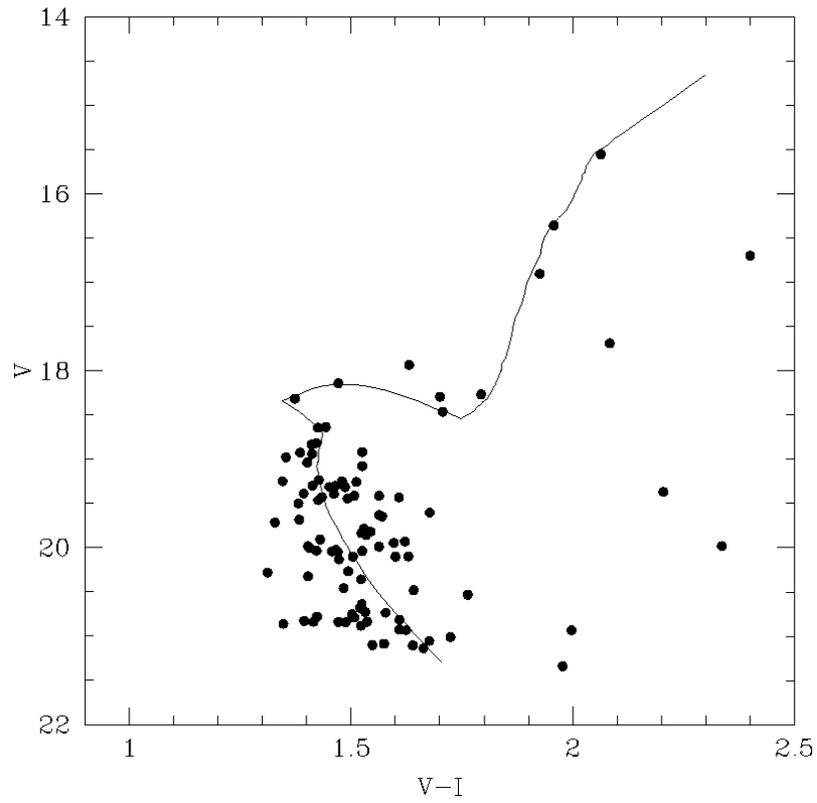}}
\caption{The adopted ($V,V-I$) diagram for Berkeley\,26. The adopted isochrone from 
Lejeune \& Schaerer (2001) are overplotted with solid lines.}
\label{fig13}
\end{figure}

\begin{figure}
\centerline{\psfig{figure=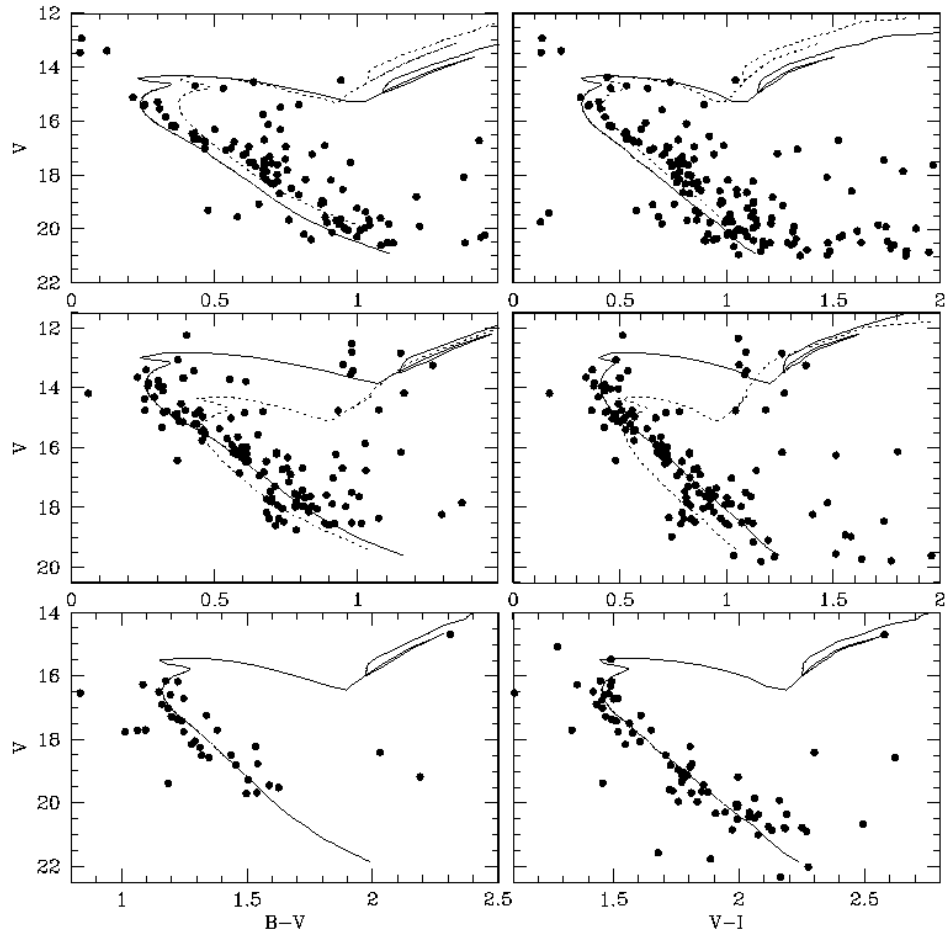}}
\caption{($V,B-V$) and ($V,V-I$) diagrams adopted for Czernik\,27 ({\it top}), Melotte\,72 
({\it middle}) and BH\,37 ({\it bottom}). {\it Czernik\,27:} the isochrones from Lejeune \& Schaerer 
(2001) according to the values published by Kim et al. \shortcite{kimetal05} and HSM08 (see Sect. 1) are overplotted with
solid and dotted lines, respectively. {\it Melotte\,72:} the isochrones from Lejeune \& Schaerer 
(2001) according to the values adopted here and those published by HSM08 (see Sect. 1) are overplotted with
solid and dotted lines, respectively. {\it BH\,37:} The adopted isochrone from 
Lejeune \& Schaerer (2001) is overplotted with solid lines.}
\label{fig14}
\end{figure}

\begin{figure}
\centerline{\psfig{figure=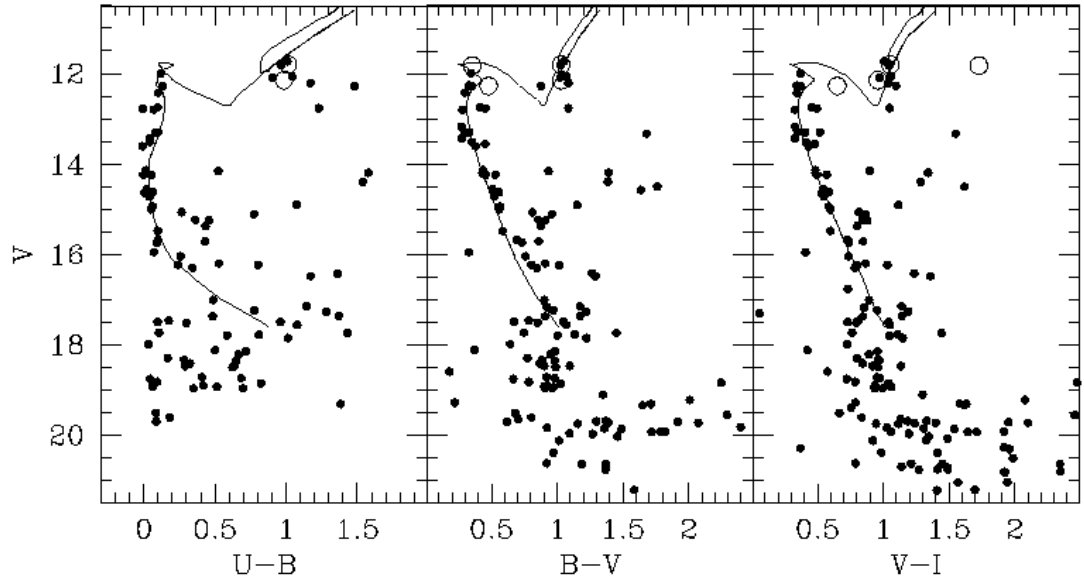}}
\caption{($V,U-B$), ($V,B-V$) and ($V,V-I$) diagrams adopted for NGC\,2479. The 
adopted isochrones from Lejeune \& Schaerer (2001) are overplotted with solid lines.
 Cluster members according to Kharchenko et al. (2005) are represented by open circles.}
\label{fig15}
\end{figure}

\end{document}